\documentclass[12pt]{article}
\pdfoutput = 1
\begin{document}
\title
{Minimal length effects in black hole thermodynamics from tunneling formalism}

\author{
{\bf {\normalsize Sunandan Gangopadhyay}$^{a,b,c}
$\thanks{sunandan.gangopadhyay@gmail.com}}\\
$^{a}$ {\normalsize Department of Physics, West Bengal State University, Barasat, Kolkata 700126, India}\\
$^{b}${\normalsize National Institute for Theoretical Physics (NITheP), Stellenbosch University, South Africa}\\
$^{c}${\normalsize Visiting Associate in Inter University Centre for Astronomy $\&$ Astrophysics (IUCAA),}\\
{\normalsize Pune, India}\\
}
\date{}

\maketitle

\begin{abstract}
{\noindent The tunneling formalism in the Hamilton-Jacobi approach is adopted to study Hawking radiation of massless Dirac particles from spherically symmetric black hole spacetimes incorporating the effects of the generalized uncertainty principle. 
The Hawking temperature is found to contain corrections from the generalized uncertainty principle. 
Further, we show from this result that the ratio of the GUP corrected energy of the particle to the GUP corrected Hawking temperature
is equal to the ratio of the corresponding uncorrected quantities.
This result is then exploited to compute the Hawking temperature 
for more general forms of the uncertainty principle having infinite number of terms. Choosing the coefficients
of the terms in the series in a specific way enables one to sum the infinite series exactly. This leads to a 
Hawking temperature for the Schwarzschild black hole that agrees with the result which accounts for the one loop back reaction effect. The entropy is finally computed and yields the area theorem upto logarithmic corrections.

}

\end{abstract}

\maketitle


\noindent The existence of a minimal observable length is a common feature 
of all candidate theories of quantum gravity such as string theory \cite{str}-\cite{kato}, loop quantum gravity \cite{garay} and noncommutative geometry \cite{nc}. The consequence of this is the modification of the Heisenberg uncertainty principle to the so called
generalized uncertainty principle (GUP)
\begin{eqnarray}
\Delta x\Delta p\geq\frac{\hbar}{2}\left\{1+ \alpha^2 (\Delta p)^2\right\}
\label{gup}
\end{eqnarray}
where $\alpha=1/(M_{p}c)=l_p/\hbar$, $M_{p}$ is the Planck mass and $l_p$ is the Planck length ($\sim 10^{-35}m$).
This idea has been considered seriously to study the properties of quantum gravity via the study of black holes. It has been shown in \cite{adler}-\cite{sg2} that an important effect of GUP on black hole thermodynamics is the existence of a remnant mass for which the evaporation stops. The semi-classical area theorem
connecting the entropy of the black hole with its horizon area is also found to get logarithmic 
corrections from the GUP.

Hawking radiation from black holes \cite{hawk1, hawk2} 
has been another area where the effect of GUP has been investigated \cite{nozari, chen}.
For instance in \cite{nozari}, incorporating the GUP effect, the tunneling approach in \cite{parikh} has been adopted to discuss the radiation of massless scalar particles from the Schwarzschild black hole. In this paper, taking into account the effects of the GUP, we investigate the tunneling of Dirac particles 
in the semi-classical approximation employing the Hamilton-Jacobi method adopted in \cite{sri}-\cite{majf}. 
The effect of GUP is incorporated by modifying the Dirac equation in curved spacetime.
The Hamilton-Jacobi method is then used to solve the equation of motion of the spinor field and then compute the imaginary part of the action for the (classically forbidden) process of $s$-wave emission across the event horizon of the black hole. The principle of ``detailed balance" is then used to relate to the Boltzmann factor for emission at the Hawking temperature.
The result is found to get corrections from the GUP parameter $\alpha$ (coupled with the energy of the emitted fermions).
We then replace the energy by the uncertainty in the position of the emitted particle near the black hole horizon
using the Heisenberg uncertainty principle and this leads to the result for the Hawking temperature obtained in
the literature from heuristic arguments \cite{sg1}. 
Interestingly, this result reveals that the ratio of the GUP corrected energy of the particle to the GUP corrected Hawking temperature
is equal to the ratio of the corresponding uncorrected quantities.  
We then apply this result to derive the Hawking temperature for
more general forms of the uncertainty principle. The results obtained are in agreement with those found in \cite{rb}.
Our results are also consistent with those (investigating the effects of GUP on the temperature and entropy of black holes)
obtained by other methods, namely, the path integral method \cite{path}, heuristic arguements involving the GUP and horizon
area of the black hole \cite{bm}, considering corrections to all orders in the Planck length to the entropy
of a scalar field on the background of the Schwarzschild black hole \cite{ykim}, computing corrections to the thermodynamics
of black holes assuming that the GUP corrected entropy-area relation is universal for all black objects \cite{mirf},
counting the degrees of freedom near the horizon of the black hole \cite{myoon}. 

To take into account the effects of quantum gravity, we adopt the generalized commutation relation 
in \cite{kem} to modify the
Dirac equation:
\begin{eqnarray}
[x_{i}, p_{j}]=i\hbar(1+\alpha^2 p^2)\delta_{ij}.
\label{comm-rel}
\end{eqnarray}
The momentum operators consistent with the above commutation relation are therefore defined as
\begin{eqnarray}
p_i=p_{0i}(1+\alpha^2 p^2)
\label{mom1}
\end{eqnarray}
where $p_{0i}$ can be interpreted as the momentum at low energies having the standard representation $p_{0i}=-i\hbar\partial_{x_{0i}}$ in
position space satisfying the standard canonical Heisenberg commutation relations ($[x_{i}, p_{0j}]=i\hbar\delta_{ij}$) and $p_i$ as the momentum at higher energies. 

\noindent The square of momentum operators upto order $\alpha^2$ is therefore given by
\begin{eqnarray}
p^2&=&p_{i}p^{i}=-\hbar^2 [1-\alpha^2 \hbar^2 \partial_{j}\partial^{j}]\partial_{i}
[1-\alpha^2 \hbar^2 \partial_{j}\partial^{j}]\partial^{i}\nonumber\\
&=&-\hbar^2 [ \partial_{i}\partial^{i}-2\alpha^2 \hbar^2  (\partial_{j}\partial^{j})( \partial_{k}\partial^{k})]
+\mathcal{O}(\alpha^4).
\label{sq}
\end{eqnarray}
The effect of the GUP also modifies the energy \cite{hos}
\begin{eqnarray}
\tilde{E}=E(1+\alpha^2 E^2)
\label{GUP_energy}
\end{eqnarray}
with $E=i\hbar\partial_{0}$ being the energy operator. The modification of the energy due to the effect of the GUP takes place at high energies and follows by translating eq.(\ref{gup}) to the energy position uncertainty bound relation \cite{land}.

We now consider a massless Dirac particle in a general
class of static, spherically symmetric spacetime of the form
\begin{eqnarray}
ds^2 =-f(r)dt^2 +\frac{1}{g(r)}dr^2 +r^2 (d\theta^2 +\sin^{2}\theta d\phi^2)
\label{metric}
\end{eqnarray}
where the horizon $r=r_H$ is given by $f(r_H)=g(r_H)=0$.

\noindent The massless Dirac equation is given by
\begin{eqnarray}
i\gamma^{\mu}\nabla_{\mu}\psi=0
\label{dirac1}
\end{eqnarray}
where the $\gamma$ matrices are defined as
\begin{eqnarray}
\gamma^{t}&=&\frac{1}{\sqrt{f(r)}}\pmatrix{i& 0\cr 0&-i\cr}~, 
~\gamma^{r}=\sqrt{g(r)}\pmatrix{0& \sigma^3\cr \sigma^3&0\cr}\nonumber\\
\gamma^{\theta}&=&\frac{1}{r}\pmatrix{0& \sigma^1\cr \sigma^1&0\cr}~,
~\gamma^{\phi}=\frac{1}{r\sin\theta}\pmatrix{0& \sigma^2\cr \sigma^2&0\cr}.
\label{gamma}
\end{eqnarray}
The covariant derivative is given by
\begin{eqnarray}
\nabla_{\mu}&=&\partial_{\mu}+\frac{i}{2}g^{\alpha\nu}{\Gamma^{\alpha}}_{\mu\nu}\Sigma_{\alpha\alpha}\nonumber\\
\Sigma_{\alpha\alpha}&=&\frac{i}{4}[\gamma_{\alpha}, \gamma_{\alpha}]~,~\{\gamma^{\mu}, \gamma^{\nu}\}=2g^{\mu\nu}.
\label{covder}
\end{eqnarray}
To get the GUP modified Dirac equation in curved spacetime, we first rewrite eq.(\ref{dirac1}) as
\begin{eqnarray}
-i\gamma^{0}\partial_{0}\psi=(i\gamma^{i}\partial_{i}-\frac{1}{2}\gamma^{\mu}g^{\alpha\nu}
{\Gamma^{\alpha}}_{\mu\nu}\Sigma_{\alpha\alpha})\psi~;~i=r,\theta,\phi.
\label{dirac2}
\end{eqnarray}
The left hand side of the above equation is related to the energy. Using
the expressions for the GUP modified energy operator (\ref{GUP_energy}) and the square of
the momentum operator (\ref{sq}), we get upto leading order in $\alpha^2$
\begin{eqnarray}
-i\gamma^{0}\partial_{0}\psi=(i\gamma^{i}\partial_{i}-\frac{1}{2}\gamma^{\mu}g^{\alpha\nu}
{\Gamma^{\alpha}}_{\mu\nu}\Sigma_{\alpha\alpha})(1-\alpha^{2}\hbar^2 \partial_{j}\partial^{j})\psi.
\label{dirac3}
\end{eqnarray}
For radial trajectories, only the ($r-t$) sector of the metric (\ref{metric}) is important and hence the above equation
can be expressed as
\begin{eqnarray}
-i\gamma^{0}\partial_{0}\psi=[i\gamma^{r}\partial_{r}-\frac{1}{2}(g^{tt}\gamma^{\mu}
{\Gamma^{r}}_{\mu t}-g^{rr}\gamma^{\mu}
{\Gamma^{t}}_{\mu r})\Sigma_{rt}](1-\alpha^{2}\hbar^2 \partial_{j}\partial^{j})\psi.
\label{dirac4}
\end{eqnarray}
Now the nonvanishing connections for the metric (\ref{metric}) are
\begin{eqnarray}
{\Gamma^{r}}_{tt}=\frac{f' g}{2}~,~{\Gamma^{t}}_{tr}=\frac{f'}{2f}~.
\label{connec}
\end{eqnarray}
To solve eq.(\ref{dirac4}), we make the following ansatz for the spin up
(i.e. positive $r$ direction)\footnote{The spin down (i.e. negative $r$ direction) case can be analysed exactly in the same way.} $\psi$:
\begin{eqnarray}
\psi(t, r)=\pmatrix{A(t, r) \cr 0\cr B(t, r)\cr0\cr}\exp[\frac{i}{\hbar}I(t, r)]
\label{sol1}
\end{eqnarray}
where $I(t, r)$ is the one particle action. Substituting this in eq.(\ref{dirac4}), we obtain the following set of equations:
\begin{eqnarray}
\label{sol2}
\frac{i}{\sqrt{f}}A\partial_{0}I +\sqrt{g}B\partial_{r}I+\alpha^{2} g^{3/2}B(\partial_{r}I)^3&=&0\\
-\frac{i}{\sqrt{f}}B\partial_{0}I +\sqrt{g}A\partial_{r}I+\alpha^{2} g^{3/2}A(\partial_{r}I)^3&=&0.
\label{sol3}
\end{eqnarray}
In the above equations, we have dropped terms of the order of $\hbar$ and terms which
do not involve the single particle action since they do not contribute to the thermodynamic entities of the black hole.

\noindent Since the metric (\ref{metric}) is stationary it has timelike Killing vectors. Hence we will look for solutions of 
this equation in the form
\begin{eqnarray}
I(t, r)=\omega_G t +W(r)
\label{sol4}
\end{eqnarray}
where $\omega_G$ is the GUP corrected energy of the particle. Substituting this in eq(s).(\ref{sol2}) and (\ref{sol3}), we get
\begin{eqnarray}
\label{sol5}
\frac{i}{\sqrt{f}}A\omega_G+\sqrt{g}B\partial_{r}W+\alpha^{2} g^{3/2}B(\partial_{r}W)^3&=&0\\
-\frac{i}{\sqrt{f}}B\omega_G+\sqrt{g}A\partial_{r}W+\alpha^{2} g^{3/2}A(\partial_{r}W)^3&=&0.
\label{sol6}
\end{eqnarray}
We shall first solve the above equations setting $\alpha^2=0$. We shall then use these solutions 
to solve the above equations (with nonzero $\alpha^2$) perturbatively upto order $\alpha^2$.
Setting $\alpha^2=0$ and noting that $\omega_G=\omega$ (the energy of the particle in the absence of GUP) 
in the $\alpha=0$ limit leads to the following solutions
\begin{eqnarray}
\label{sol7a}
A&=&iB~,~\partial_{r}W(r)=\frac{\omega}{\sqrt{f(r)g(r)}}\\
A&=&-iB~,~\partial_{r}W(r)=-\frac{\omega}{\sqrt{f(r)g(r)}}~.
\label{sol7}
\end{eqnarray}
Substituting the value of $A$ in eq(s).(\ref{sol5}) and (\ref{sol6}) and $\partial_{r}W(r)$ in the third term of eq(s).(\ref{sol5}) and (\ref{sol6}) leads to the following equation for $W(r)$ upto order $\alpha^2$:
\begin{eqnarray}
\partial_{r}W(r)&=&\pm\frac{\omega_G}{\sqrt{f(r)g(r)}}\left\{1-\frac{\alpha^2 \omega^3}{f(r)\omega_G}\right\}\nonumber\\
&\approx&\pm\frac{\omega_G}{\sqrt{f(r)g(r)}}\left\{1-\frac{\alpha^2 \omega^2}{f(r)}\right\}
\label{sol8}
\end{eqnarray}
where we have set $\omega_G \approx\omega$ in the second term of the above equation since the computations are upto order $\alpha^2$.
Solving this equation, we obtain
\begin{eqnarray}
W(r)=\pm\omega_G\int^{r}_{0}\frac{dr}{\sqrt{f(r)g(r)}}\left\{1-\frac{\alpha^2 \omega^2}{f(r)}\right\}.
\label{sol9}
\end{eqnarray}
The limits of the integration are chosen such that the particle goes through the horizon $r=r_{H}$.
Hence the solution for $I(t, r)$ reads
\begin{eqnarray}
I(t, r)=\omega_G t \pm\omega_G\int^{r}_{0}\frac{dr}{\sqrt{f(r)g(r)}}\left\{1-\frac{\alpha^2 \omega^2}{f(r)}\right\}.
\label{sol10}
\end{eqnarray}
Using the above solution and eq.(\ref{sol1}), the ingoing and outgoing solutions
of the Dirac equation (\ref{dirac4}) can be written as
\begin{eqnarray}
\label{in}
\psi_{in}&\sim&\exp\left[\frac{i}{\hbar}\left(\omega_G t +\omega_G\int^{r}_{0}\frac{dr}{\sqrt{f(r)g(r)}}\left\{1-\frac{\alpha^2 \omega^2}{f(r)}\right\}\right)\right]\\
\psi_{out}&\sim&\exp\left[\frac{i}{\hbar}\left(\omega_G t -\omega_G\int^{r}_{0}\frac{dr}{\sqrt{f(r)g(r)}}\left\{1-\frac{\alpha^2 \omega^2}{f(r)}\right\}\right)\right].
\label{out}
\end{eqnarray}
Now the nature of the coordinates change for the tunneling of a particle across the horizon of the black hole. 
The sign of the metric coefficients in the ($r$-$t$) sector gets altered. 
This indicates that the $t$-coordinate has an imaginary part for the
crossing of the horizon of the black hole and correspondingly a temporal contribution will be there to the probabilities
for the ingoing and outgoing particles.

\noindent Hence, the ingoing and outgoing probabilities of the particle are given by
\begin{eqnarray}
\label{in1}
P_{in}&=&|\psi_{in}|^2\sim\exp\left[\frac{2}{\hbar}\left(\omega_G Im~t +\omega_G Im \int^{r}_{0}\frac{dr}{\sqrt{f(r)g(r)}}\left\{1-\frac{\alpha^2 \omega^2}{f(r)}\right\}\right)\right]\\
P_{out}&=&|\psi_{out}|^2\sim\exp\left[\frac{2}{\hbar}\left(\omega_G Im~t -\omega_G Im \int^{r}_{0}\frac{dr}{\sqrt{f(r)g(r)}}\left\{1-\frac{\alpha^2 \omega^2}{f(r)}\right\}\right)\right]
\label{out1}
\end{eqnarray}
where $Im$ denotes the imaginary part of the expression. Now the ingoing probability $P_{in}$ has to be unity in the
classical limit ($\hbar\rightarrow0$), that is when there is no reflection and everything is absorbed.
Therefore, in the classical limit, eq.(\ref{in1}) leads to
\begin{eqnarray}
\label{in2}
Im~t =-Im \int^{r}_{0}\frac{dr}{\sqrt{f(r)g(r)}}\left\{1-\frac{\alpha^2 \omega^2}{f(r)}\right\}.
\end{eqnarray}
Using this, the probability of the outgoing particle therefore reads
\begin{eqnarray}
P_{out}&=&|\psi_{out}|^2\sim\exp\left[-\frac{4}{\hbar}\omega_G Im \int^{r}_{0}\frac{dr}{\sqrt{f(r)g(r)}}
\left\{1-\frac{\alpha^2 \omega^2}{f(r)}\right\}\right].
\label{out2}
\end{eqnarray}
Now we apply the principle of ``detailed balance" \cite{sri} which states that in a system with a temperature $\beta^{-1}$, the emission
and absorption probabilities are related by $P_{em}=e^{-\beta E}P_{in}$ which in this case (for a particle with energy $\omega_G$
emitted at a Hawking temperature $T_h$) reads
\begin{eqnarray}
P_{out}=\exp\left(-\frac{\omega_G}{T_{h}}\right).
\label{db}
\end{eqnarray}
From eq(s).(\ref{out2}, \ref{db}), we obtain the GUP corrected temperature of the black hole to be
\begin{eqnarray}
T_h &=& \frac{\hbar}{4}\left(Im \int^{r}_{0}\frac{dr}{\sqrt{f(r)g(r)}}\left\{1-\frac{\alpha^2 \omega^2}{f(r)}\right\}\right)^{-1}
\nonumber\\
&=&\frac{\hbar}{4}\left(Im \int^{r}_{0}\frac{dr}{f(r)}\left\{1-\frac{\alpha^2 \omega^2}{f(r)}\right\}\right)^{-1}~;~for~
f(r)=g(r).
\label{bhtemp}
\end{eqnarray}
We shall now compute this for the  Schwarzschild black hole for which the metric coefficients are
\begin{eqnarray}
f(r)=g(r)=\left(1-\frac{r_H}{r}\right)~;~r_H =2M.
\label{schbh}
\end{eqnarray}
Setting $r-2M=\varepsilon e^{i\theta}$ in eq.(\ref{bhtemp}) and taking the limit $\varepsilon\rightarrow0$ at the end of the
calculation yields
\begin{eqnarray}
T_h &=& \frac{\hbar}{4}\left(Im \lim_{\varepsilon\rightarrow0}\int_{\pi}^{2\pi}\frac{i\varepsilon e^{i\theta}d\theta}
{\varepsilon e^{i\theta}}(2M+\varepsilon e^{i\theta})\left\{1-\frac{\alpha^2 \omega^2}{\varepsilon e^{i\theta}}(2M+\varepsilon e^{i\theta})\right\}\right)^{-1}\nonumber\\
&=& \frac{T_H}{(1-\alpha^2 \omega^2)}\nonumber\\
&=&T_H (1+\alpha^2 \omega^2)+\mathcal{O}(\alpha^4)~;~T_H =\frac{\hbar}{8\pi M}~.
\label{t1}
\end{eqnarray}
Now using the saturated form of the uncertainty principle $\omega \Delta x =\hbar/2$ \cite{land}
(which follows from the saturated form of the Heisenberg uncertainty principle 
$\Delta x\Delta p=\hbar/2$) in the above equation, we get
\begin{eqnarray}
T_h =T_H \left[1+\frac{\alpha^2 \hbar^2}{4(\Delta x)^2}\right]+\mathcal{O}(\alpha^4)
\label{t2}
\end{eqnarray}
which agrees with earlier finding \cite{sg1} where the result has been derived from heuristic arguments.
Note that $\Delta x$ is the uncertainty in the position of the particle near the black hole horizon and is of the order
of the horizon radius $r_H$ \cite{adler}. Further this uncertainty in the position of the particle vanishes when it is
far away from the horizon of the black hole.

\noindent It is now interesting to note that eq.(\ref{t1}) can be put in the following form (by multiplying it
by the inverse of the energy of the particle $\omega$)
\begin{eqnarray}
\frac{\omega_G}{T_{h}}=\frac{\omega}{T_{H}}
\label{prop}
\end{eqnarray}
where 
\begin{eqnarray}
\omega_{G}&=&\omega\left\{1+\alpha^2 \omega^{2}\right\}
\label{G_energy}
\end{eqnarray}
is the GUP corrected energy of the particle and is consistent with eq.(\ref{GUP_energy}). Note that eq.(\ref{prop}) reveals
that the ratio of the GUP corrected energy of the particle to the GUP corrected Hawking temperature
is equal to the ratio of the corresponding uncorrected quantities.

\noindent We shall now use eq.(\ref{prop}) (to compute the GUP corrected Hawking temperature of the black hole) for more general forms of the uncertainty principle 
\begin{eqnarray}
\omega_{G}=\omega\left\{1+a_1 \left(\frac{\hbar}{2\Delta x}\right)^{2}+a_2 \left(\frac{\hbar}{2\Delta x}\right)^{4} +...\right\}.
\label{G_energy}
\end{eqnarray}
Substituting this in eq.(\ref{prop}) gives
\begin{eqnarray}
T_{h}=T_{H}\left\{1+a_1 \left(\frac{\hbar}{2\Delta x}\right)^{2}+a_2 \left(\frac{\hbar}{2\Delta x}\right)^{4} +...\right\}.
\label{G_temp}
\end{eqnarray}
Setting $a_n =\alpha^{2n}$, $n=1, 2, 3, ...$, the infinite series on the right hand side of the above equation
can be summed up and yields
\begin{eqnarray}
T_{h}=\frac{T_{H}}{\left[1-\frac{\alpha^2 \hbar^2}{4(\Delta x)^2}\right]}~.
\label{G_temp1}
\end{eqnarray}
Note that the above expression for the GUP corrected Hawking temperature is an exact result which can be applied to spherically symmetric black hole spacetimes.
The expression also agrees with eq.(\ref{t2}) upto order $\alpha^2$.

We shall now apply this result to compute the GUP corrected Hawking temeperature for the Schwarzschild black hole.
Near the horizon of the Schwarzschild black hole, the position uncertainty of a particle will be of the order of the Schwarchild radius
of the black hole \cite{adler},\cite{medved}
\begin{eqnarray}
\Delta x=\varepsilon r_{s}~;~
r_{s}=2M
\label{e2}
\end{eqnarray}
where $\varepsilon$ is a calibration factor and $r_s$ is the Schwarzschild radius. 
Substituting this relation in eq.(\ref{G_temp1}), we obtain
\begin{eqnarray}
T_{h}=\frac{T_{H}}{\left[1-\frac{\alpha^2 \hbar^2}{16\varepsilon^2 M^2}\right]}~.
\label{G_temp2}
\end{eqnarray}
Remarkably the above result agrees with the one involving the one loop back reaction effect on the spacetime \cite{fur}.

\noindent We now determine the black hole entropy from the first law of black hole thermodynamics given by
\begin{eqnarray}
S=\int \frac{dM}{T_h}~.
\label{ent}
\end{eqnarray}
Using eq.(\ref{G_temp2}) and performing the above integration leads to
\begin{eqnarray}
S&=&4\pi M^2 -\frac{\alpha^2 \pi}{2\varepsilon^2}\ln M +constant\nonumber\\
&=&\frac{A}{4}-\frac{\alpha^2 \pi}{4\varepsilon^2}\ln \left(\frac{A}{16\pi}\right) +constant
\label{en1}
\end{eqnarray}
where $A=4\pi r_{s}^2 =16\pi M^2$ is the area of the horizon. The above result is the well known area theorem with 
logarithmic corrections arising from the GUP. 

We now summarize our findings. In this paper, based on the Hamilton-Jacobi approach to tunneling, we
compute the GUP corrected Hawking temperature due to tunneling of massless Dirac particles for
spherically symmetric black hole spacetimes. The method consists in first deriving the GUP modified
Dirac equation in the black hole spacetime and then applying the tunneling formalism. 
The Hawking temperature is found to contain corrections from the GUP. The result is also consistent with \cite{sg1}
when the energy of the emitted particles is replaced by the uncertainty in the position of the particle 
(near the black hole horizon) using the Heisenberg uncertainty principle. Further, it follows from the GUP
corrected Hawking temperature that the ratio of the GUP corrected energy of the particle to the GUP corrected Hawking temperature
is equal to the ratio of the corresponding uncorrected quantities.
We then apply this result to calculate the Hawking temperature 
for more general forms of the uncertainty principle having infinite number of terms. Setting the coefficients
of the terms in the series in a specific way allows one to sum the infinite series exactly. Interestingly, this leads to a 
Hawking temperature for the Schwarzschild black hole that agrees with the result accounting for the one loop back reaction effect. The entropy is finally computed from the first law of black hole thermodynamics and yields the area theorem upto logarithmic corrections.
It is to be noted that the presence of remnants evident in our approach (from the expression of the GUP corrected Hawking temperature of the black hole (\ref{t2}, \ref{G_temp1})) \cite{sg1, sg2} is consistent with that present in the framework
of rainbow gravity \cite{khal}-\cite{yling}.



 \end{document}